\documentclass[sigconf,natbib=true]{acmart}
\AtBeginDocument{%
  \providecommand\BibTeX{{%
    \normalfont B\kern-0.5em{\scshape i\kern-0.25em b}\kern-0.8em\TeX}}}

\copyrightyear{2022}
\acmYear{2022}
\setcopyright{acmcopyright}\acmConference[CIKM '22]{Proceedings of the 31st ACM International Conference on Information and Knowledge Management}{October 17--21, 2022}{Atlanta, GA, USA}
\acmBooktitle{Proceedings of the 31st ACM International Conference on Information and Knowledge Management (CIKM '22), October 17--21, 2022, Atlanta, GA, USA}
\acmPrice{15.00}
\acmDOI{10.1145/3511808.3557244}
\acmISBN{978-1-4503-9236-5/22/10}

\newcommand{\rlmp}{RMS}
\newcommand{\rec}{HRec}
\newcommand{\rlrec}{RMS-HRec}

\newcommand{\stitle}[1]{\vspace*{0.4em}\noindent{\bf #1:\/}}

\usepackage{multirow}
\usepackage{graphicx}
\usepackage{float}

\usepackage{subfig}

\begin{document}

\title{Automatic Meta-Path Discovery for Effective Graph-Based Recommendation}


\author{Wentao Ning}
\affiliation{
    \institution{The University of Hong Kong}
    \institution{Southern University of Science and Technology}
    \country{Hong Kong and Shenzhen, China}
}
\email{wtning@cs.hku.hk}

\author{Reynold Cheng}
\authornote{Reynold Cheng is a corresponding author. He is also affiliated with Musketeers Foundation Institute of Data Science, HKU and Guangdong–Hong Kong-Macau Joint Laboratory.}
\affiliation{
    \institution{The University of Hong Kong}
    \country{Hong Kong, China}
}
\email{ckcheng@cs.hku.hk}

\author{Jiajun Shen}
\affiliation{
    \institution{TCL Research}
    \country{Hong Kong, China}
}
\email{sjj@tcl.com}

\author{Nur Al Hasan Haldar}
\affiliation{
    \institution{The University of Western Australia}
    \country{Perth, Australia}
}
\email{nur.haldar@uwa.edu.au}

\author{Ben Kao}
\affiliation{
    \institution{The University of Hong Kong}
    \country{Hong Kong, China}
}
\email{kao@cs.hku.hk}

\author{Xiao Yan}
\affiliation{
    \institution{Southern University of Science and Technology}
    \country{Shenzhen, China}
}
\email{yanx@sustech.edu.cn}

\author{Nan Huo}
\affiliation{
    \institution{The University of Hong Kong}
    \country{Hong Kong, China}
}
\email{huonan@connect.hku.hk}

\author{Wai Kit Lam}
\author{Tian Li}
\affiliation{
    \institution{TCL Research}
    \country{Hong Kong, China}
}
\email{tian23.li@tcl.com}


\author{Bo Tang}
\authornote{Bo Tang is a corresponding author. He is also affiliated with Research Institute of Trustworthy Autonomous Systems and Guangdong Provincial Key Laboratory of Brain-inspired Intelligent Computation, Shenzhen, China.}
\affiliation{
    \institution{Southern University of Science and Technology}
    \country{Shenzhen, China}
}
\email{tangb3@sustech.edu.cn}

\renewcommand{\shortauthors}{Wentao Ning et al.}




\begin{abstract} 

    \textit{Heterogeneous Information Networks} (HINs) are labeled graphs that depict relationships among different types of entities (e.g., users, movies and directors). For HINs, \textit{meta-path-based recommenders} (MPRs) utilize meta-paths (i.e., abstract paths consisting of node and link types) to predict user preference, and have attracted a lot of attention due to their explainability and performance. We observe that the performance of MPRs is highly sensitive to the meta-paths they use, but existing works manually select the meta-paths from many possible ones. Thus, to discover effective meta-paths automatically, we propose the \textit{Reinforcement learning-based Meta-path Selection} (\rlmp) framework. Specifically, we define a vector encoding for meta-paths and design a policy network to extend meta-paths. The policy network is trained based on the results of downstream recommendation tasks and an early stopping approximation strategy is proposed to speed up training. \rlmp\ is a general model, and it can work with all existing MPRs. We also propose a new MPR called \rlrec, which uses an attention mechanism to aggregate information from the meta-paths. We conduct extensive experiments on real datasets. Compared with the manually selected meta-paths, the meta-paths identified by \rlmp~consistently improve recommendation quality. Moreover, \rlrec \ outperforms state-of-the-art recommender systems by an average of 7\% in hit ratio. The codes and datasets are available on https://github.com/Stevenn9981/RMS-HRec.

\end{abstract}




\maketitle


\section{Introduction}\label{sec:intro}

\begin{figure}[t]
	\centering
	\includegraphics[width=7cm]{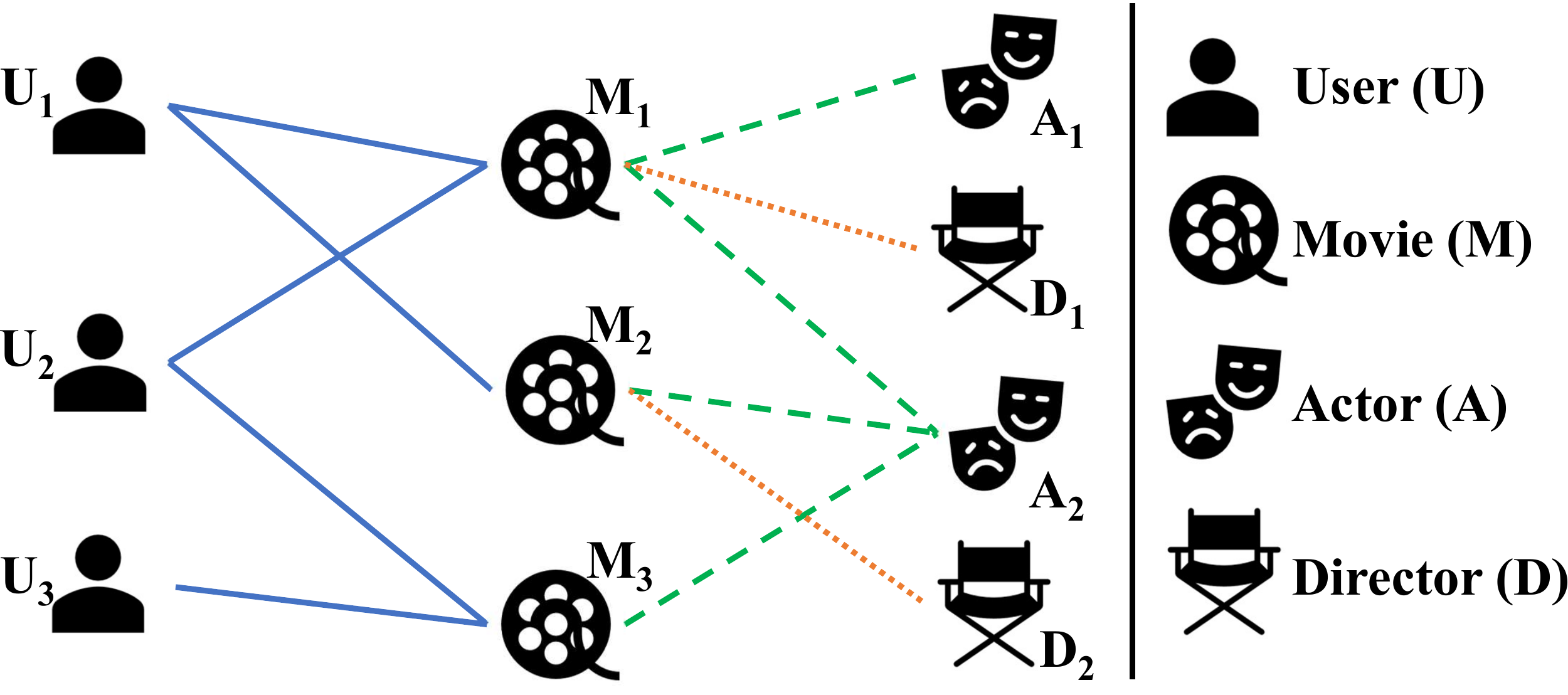}
	\caption{An example of a movie HIN. It contains four node types (i.e., user, movie, actor, and director) and three relationship types (i.e., watched, acted, directed) among the nodes.}
	\label{fig:hin}
	\vspace{-1em}
\end{figure}

A Heterogeneous Information Network (HIN)~\cite{huang2016meta, ZhangYLXM16, HuangZDWC18, WangZXG18} consists of multiple types of entities and links (or relations), and Figure~\ref{fig:hin} illustrates an example of movie HINs. Due to its ability to model the complex interactions among different entities, HINs are widely used in fields such as natural language processing~\cite{wang2020heterogeneous,zheng-etal-2020}, community detection~\cite{fang2020effective, tu2019a}, and bioinformatics~\cite{ZhaoHYWS22, KumarTMG22}. HINs have also been used to support recommender systems. HINs provide more information than user-item interaction logs commonly utilized by recommenders, and have attracted plenty of attention recently~\cite{DongCS17,AAAI21JiZWSWTLH, AiACZ18}.

Meta-path, which consists of a sequence of relations between different node types~\cite{SunHYYW11}, is often utilized to extract information from HINs for recommendation. For example, in Figure \ref{fig:hin}, $Movie \rightarrow Actor \rightarrow Movie$ (abbreviated as $MAM$) is a meta-path and can be adopted to find movies having the same actor. A few \textit{meta-path}-based recommenders (MPRs) have been proposed in recent years~\cite{YuRSSKGNH13, MM21HaoX0H, AAAI21JiZWSWTLH}. For example, HERec~\cite{herec2018shi} collects random walk samples according to meta-paths to train user and item embeddings. MCRec~\cite{hu2018leveraging} utilizes a convolutional layer to train meta-path representations based on their sampled path instances. MEIRec~\cite{FanZHSHML19} utilizes meta-paths to select related node neighbors and fuses information from them to obtain node embeddings. These MPRs achieved better performance compared with existing recommenders, and are favored as they  \textit{can also provide the explainability}~\cite{guo2020survey, fu2020magnn, HuangFQSLX19, Sun00BHX18} to the recommendation results. For example, MCRec~\cite{hu2018leveraging} learns the weights of meta-paths for each user-item pair, if the weight of meta-path $UMAM$ (refer to  Figure~\ref{fig:hin}) is the largest among all the meta-paths for a user-item pair in the recommendation results, the reason that the specific movie is recommended to the user is more likely because he has ever seen many movies starring an actor who also stars in this movie.

Existing MPRs require a set of meta-paths as input. Our experiments reveal that the performance of MPRs is highly sensitive to the choice of meta-paths. For example, on the Yelp dataset (Section~\ref{sec:exp}), the use of different meta-path sets can result in a significant
performance difference in recommendation. For example, Table~\ref{tab:mp_exp} shows that meta-path set 2 yields a recommendation quality gain of $17.7\%$ over meta-paths set 1.


Although using the right set of meta-paths is crucial to the performance of MPRs, the issues of finding an effective meta-path set for recommendation have not been well addressed in the literature. Existing MPR works often claim that domain experts' knowledge can be used to obtain meta-paths. The fact is that selecting correct meta-paths for recommendation is very challenging. First, due to the complexity of an HIN, the search for the optimal meta-paths is highly expensive -- the complexity of finding $m$ length-$l$ meta-paths for $n$ node types is exponential (i.e., $O(n^{lm})$).  Second, our experiments show that for a given HIN, the best meta-path set varies across different MPRs; a meta-path set that is the best for an MPR may not work well for another MPR. Third, the structures of the meta-paths required by different MPRs can vary. While some recommendation models need meta-paths that start with a user type and end with an item type~\cite{hu2018leveraging,FanZHSHML19}, others need meta-paths that both start and end with a user or item type~\cite{herec2018shi, peagnn2020han}. These observations call for better methods for discovering meta-paths automatically.

\begin{table}[]
	\centering
	\begin{tabular}{|c|c|c|c|}
	\hline
	Set ID & Input meta-path set & HR@1   & Improv.                \\ \hline
	1 & UUU, BUB      & 0.0413 & \multirow{2}{*}{17.7\%} \\ \cline{1-3}
	2 & UBU, BUB      & 0.0486 &                        \\ \hline
	\end{tabular}
	\caption{Recommendation quality based on different meta-path sets on Yelp. Here, `U' represents `User', `B' means `Business', and `HR@1' is a recommendation quality metric. 
	}
	\label{tab:mp_exp}
	\vspace{-3em}
	\end{table}

To address these issues, we propose a model-independent Reinforcement learning (RL)-based Meta-path Selection (\rlmp) framework to figure out the meaningful meta-paths for recommendation and it is the first framework that can be directly applied to existing MPRs.
We train a policy network (agent) to automatically generate a relation to extend the current meta-path set to take the place of the human intervention in the designing of meta-path sets and use the encoding of the current meta-path set as input. 
We optimize the meta-path selection policy network by exploiting the performance gain of the newly generated meta-path set from the downstream recommender. 
In addition, to incorporate the \textit{meta-path} selection in the recommendation tasks, we also propose a new MPR dubbed \rec \ and obtain an advanced algorithm \rlrec \ by integrating \rec \ with \rlmp. 
\rec \ generates user/item meta-path subgraphs via the generated meta-paths and then obtains the embeddings via a graph attention network. Then another attention mechanism is utilized to fuse the user/item embeddings from each meta-path.

Specifically, the above-mentioned procedures are non-trivial due to the following challenges. 1) It is arduous to get the performance gain of each meta-path set from a large number of possible meta-path sets because it will result in a considerable time consumption to train the RL model.  2) The huge training time and memory consumption of \rec \ make direct training impossible on the real-world graphs. To deal with these challenges, inspired by the hyperparameter optimization techniques~\cite{ICML20StandleyZCGMS20, IJCAI15DomhanSH15, JMLR17LiJDRT17},
we propose an early stopping approximation technique, namely testing meta-path sets on lightly trained models, to ease the training time burden (Challenge 1, details in Section \ref{subsec: rms_training}).
Additionally, we also propose some strategies to control the density of generated meta-path subgraphs to avoid large GPU memory usage for more effective training of \rec \ (Challenge 2, details in Section \ref{sec: hrec}). 
Our experiments show the meta-path set selected by \rlmp \ can improve accuracy by more than 20\% in some cases compared to other strategies, and \rlmp \ can find some meta-paths that are effective but not obvious. Also, \rlrec \ performs better than other state-of-the-art recommenders within a comparable training overhead.

The main contributions of this work are as below:
\begin{itemize}
	\item We propose a model-independent meta-path selection framework \rlmp, which can be plugged into any meta-path-based recommendation model, and an early stopping strategy to make the training more efficient. 
	\item Equipped with \rlmp, we develop a new meta-path-based recommender \rlrec \ and design training strategies to make the training process more efficient and scalable.
	\item We conduct extensive experiments to evaluate \rlmp~ and \rlrec. Experimental results demonstrate the performance of a recommender is very sensitive to the input meta-paths and our proposed models can discover important meta-paths that are easily neglected during manual designing.
\end{itemize}

The rest of the paper is organized as follows. Section~\ref{sec:pr} gives some basic definitions. Section~\ref{sec: rms} elaborates our meta-path selection framework \rlmp~and Section~\ref{sec: hrec} illustrates our proposed recommender \rec. Section~\ref{sec:exp} proves the effectiveness of our methods by extensive experiments. Section ~\ref{sec:rel} describes the related work.  Section~\ref{sec:con} draws a conclusion.

\section{Preliminary} \label{sec:pr}

In this section, we introduce some necessary definitions~\cite{SunHYYW11, han2019}.

\begin{definition}
	(\textbf{Heterogeneous Information Network}) A Heterogeneous Information Network (HIN) is a  graph, denoted as $\mathcal{G} = (\mathcal{V}, \mathcal{E}, \mathcal{N}, \mathcal{R})$, with multiple types of nodes and edges.  $\mathcal{V}$ represents the node set, which is associated with a node type mapping function $\zeta: \mathcal{V} \rightarrow \mathcal{N}$, where $\mathcal{N}$ is the node type set. 
	$\mathcal{E}$ denotes the edge set, which is also associated with an edge type mapping function $\psi: \mathcal{E} \rightarrow \mathcal{R}$, where $\mathcal{R}$ denotes the relation type set. 
\end{definition}

\stitle{Example} 
Figure \ref{fig:hin} shows an example of HIN, which consists of four types of nodes (i.e., $ \mathcal{N}$ = \{user, movie, actor, director\}) and three types of relations/edges (i.e., $\mathcal{R}$ = \{watch/watched, act/acted, direct/directed\}).

\begin{definition}
		(\textbf{Meta-path}) Given an HIN $\mathcal{G}$, a meta-path $\mathcal{M}$ is a sequence of node types  in the form of $n_1 \rightarrow n_2 \rightarrow \cdots \rightarrow n_{l}$ (abbreviated as $n_1n_2\cdots n_{l}$), where $n_j \in \mathcal{N}$ denotes the node type.
\end{definition}


\stitle{Example} $User \rightarrow  Movie \rightarrow Actor \rightarrow Movie$  (abbreviated as $UMAM$) is a meta-path in Figure  \ref{fig:hin}.

\begin{definition}
	(\textbf{Meta-path Instances}) Given an HIN $\mathcal{G}$ and a meta-path $\mathcal{M}$, a meta-path instance $I^\mathcal{M}_i$ is defined as a node sequence in $\mathcal{G}$ following the meta-path $\mathcal{M}$. 
\end{definition}

\stitle{Example} Suppose meta-path $\mathcal{M} = UMAM$ and $\mathcal{G}$ is the HIN shown in Figure \ref{fig:hin}, then one meta-path instance $I^\mathcal{M}_i$ in $\mathcal{G}$ can be $U_1  M_1 A_2 M_3$.

\begin{definition}
	(\textbf{Meta-path Neighbors}) Given an HIN $\mathcal{G}$ and a meta-path $\mathcal{M}$, the meta-path neighbors of node $n$ are a set of nodes $N^\mathcal{M}_n$ in $\mathcal{G}$ which is connected to $n$ via the meta-path $\mathcal{M}$. 
\end{definition}

\stitle{Example} As shown in Figure \ref{fig:hin},  suppose meta-path $\mathcal{M} = MAM$, then node $M_1$'s meta-path neighbors $N^\mathcal{M}_{M_1} = \{ M_2, M_3\}$. 

\begin{figure}[t]
	\centering
	\includegraphics[width=7.5cm]{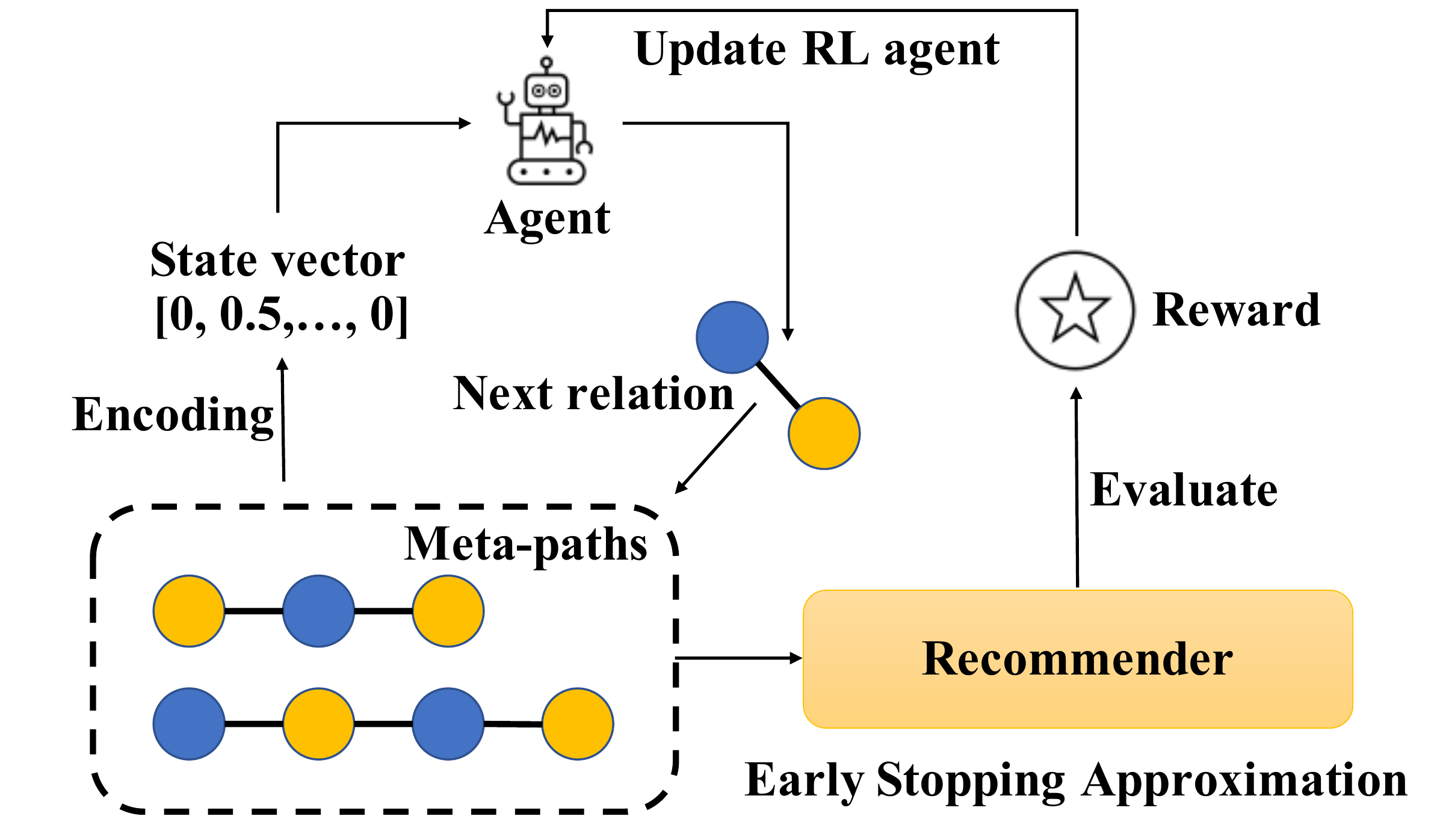}
	\caption{An overview of \rlmp~framework. \rlmp \ uses an agent (a policy network) to generate new meta-path sets based on the encoding of the current meta-path set. Then the downstream recommender uses the newly generated meta-path set as input and its performance gain is used to update the agent. For the recommender, it can be any meta-path-based model. We also propose an early stopping approximation technique to speed up the training process.}
	\label{fig:overview}
	\vspace{-1em}
\end{figure}

\section{\rlmp~for Meta-path Selection}
\label{sec: rms}

In this section, we first discuss the overview workflow of \rlmp, then we elaborate on the core components of our \rlmp \ model in detail. In the end, we illustrate how we train the model and describe our proposed strategy, which is used to speed up the training process.

\subsection{Overview}
\label{subsec: rms_overview}
We propose an RL-based framework \rlmp \ to select \textit{meta-paths} for the MPRs. The overview of \rlmp \ is shown in Figure~\ref{fig:overview}. The RL agent is a policy network that aims to investigate high-quality meta-paths. It takes the encoding of the current meta-path set as input and predicts the next relation to extend the current meta-path set. The reward will be the performance gain of the downstream recommender after using the new meta-path set and used to update the agent. Note that we also propose an early stopping approximation technique that makes the training process much more efficient. 


\subsection{Main Components of \rlmp}
\label{subsec: rms_comp}

Reinforcement Learning (RL) follows a Markov Decision Process (MDP)~\cite{SuttonB98} formulation, which contains  a set of states $\mathcal{S}$,  a set of actions $\mathcal{A}$, a decision policy $\mathcal{P}$, and a reward function $\mathcal{R}$. 
To maximize the cumulative rewards, the agent learns to consider actions based on the current state in the environment derived from an HIN. We design the key components of \rlmp \ and formalize it with the tuple $(\mathcal{S}, \mathcal{A}, \mathcal{P}, \mathcal{R})$. 
The procedure is formulated as below:

	\textbf{State ($\mathcal{S}$):} 
	It is important to store the structural information of current meta-paths to guarantee an agent can take an effective action based on the current state. We assign an ID to each relation type on an HIN (from 1 to $n$, where $n$ is the number of relation types).
	We encode each meta-path with a vector of length $n$, and each entity is the number of the corresponding relation type in this meta-path. It can make sure all the encodings have the same length.
	For instance, if there are 6 types of relations on an HIN\footnote{Note that relations with opposite head and tail nodes should be treated as two different relations, e.g. $U-M$ and $M-U$ are two different relation types.}, we will assign the relation type IDs to $1...6$. If the ID representation of a meta-path $\phi$ is $[2, 6, 6, 4]$, then its encoding $E_\phi$ will be $(0, 1, 0, 1, 0, 2)$. 
	
	The state $s_i$ at step $i$ is represented by the embedding of the meta-path set $\Phi_i$ at step $i$. 
	To embed a meta-path set, we add up all of the encodings of meta-paths in the meta-path set and apply an L2 normalization as below:
	 \begin{displaymath}
		\label{eq: state}
		s_i = Normalize( \sum_{\phi \in \Phi_i} E_{\phi} ),
	\end{displaymath}       
	where $\Phi_i$ denotes the meta-path set at step $i$, $E_\phi$ represents the encoding of meta-path $\phi$. This encoding method can reflect the weights of each relation in a meta-path set effectively.
	
	Note that different algorithms may require different meta-path forms. Some models may need the meta-paths starting and ending with a user or item node type. Then, we use $\{User-Item-User\}$ or $\{Item-User-Item\}$ as the initial meta-path set. If the algorithms require the meta-paths that strictly start with a user type node and end with an item type node, then we set $\{User-Item\}$ as the initial meta-path set. The initial meta-path set contains only one meta-path that directly connects users/items and reflects the user-item interactions since it is crucial for recommendation. 
	
	\textbf{Action ($\mathcal{A}$):} The action space $\mathcal{A}_{s_i}$ for a state $s_i$ is all the relations $(r_1, r_2, ... )$ that appear on an HIN, with a special action $STOP (r_0)$. At each step, the policy network will predict an action to extend the current meta-path set to yield a higher reward. Here, if the policy network selects the action $STOP$ or the current step exceeds the maximal step limit $I$, the meta-paths will not be extended. 
	
	If the predicted action is a relation on the HIN, we concatenate it with a complementary relation to make it a symmetric meta-path.  This is done to ensure that the extended meta-paths will also start and end with a user/item. 
	Then we try to extend all the current meta-paths with this symmetric meta-path and also insert this symmetric meta-path into the current meta-path set. Note that the previous meta-paths will also be copied and preserved in the meta-path set and will not be removed to make sure important information will not be overlooked.

	For instance, if we want a meta-path set that starts and ends with a user type node, in a movie HIN (see Figure~\ref{fig:hin}), the current set is $\{U-M-U\}$ and the predicted action (relation) is $M-A$, then we will auto-complete it with the relation $A-M$ and it will become $M-A-M$. Then we traverse each meta-path in the current meta-path set to find the first positions where we can insert this relation into each meta-path. The meta-path set at the next step will be $\{U-M-U, U-M-A-M-U\}$. However, $M-A-M$ does not start with a user type, so it will not be added to the meta-path set. This strategy ensures the generated meta-paths meet the requirements of the recommender. 
	
	\textbf{Policy ($\mathcal{P}$):}  The decision policy $\mathcal{P}$, is a state transition function that searches for the effective action based on the current state. The goal of $\mathcal{P}$ is to train a policy that tries to maximize the discounted cumulative reward $R = \sum_{i=i_{0}}^{I} \gamma^{i-i_{0}} r_{i}$, where $I$ is the maximal step limit and $\gamma$ is a real number between 0 and 1. $\gamma$ is acted as a discount factor to make the far-future reward less important than a near-future reward.
	
	
\begin{figure*}[]
	\centering
	\includegraphics[width=0.95\textwidth]{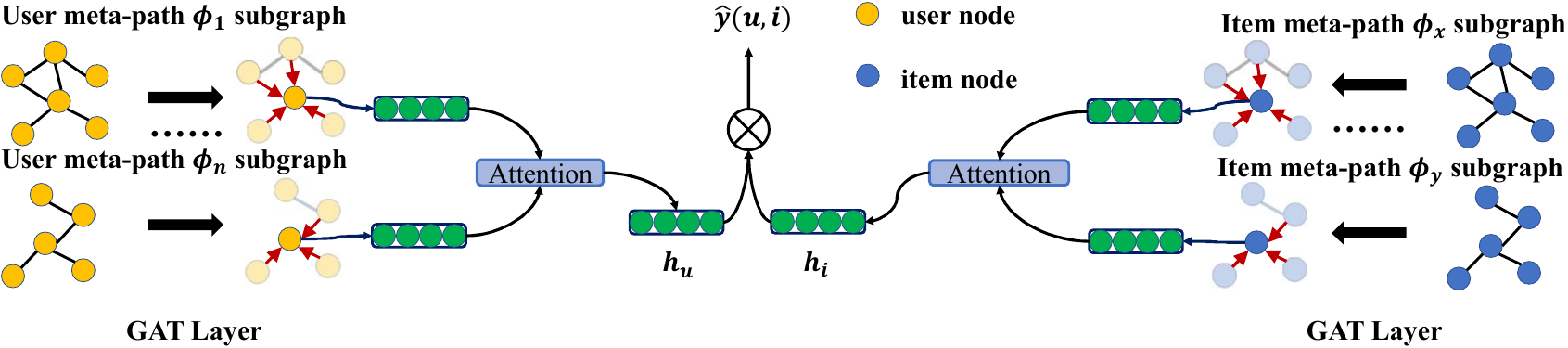}
	\caption{An overview of \rec. We generate a homogeneous meta-path subgraph for each meta-path based on meta-path connectivity, then a GAT layer~\cite{GATvelivckovic2018graph}  is used in each subgraph to learn the node embeddings by aggregating the neighbors' embedding. After that, another attention mechanism is proposed to fuse the embeddings from each meta-path to get a better representation of nodes. In the end, we calculate the recommendation scores and train the model.}
	\label{fig:recommender}
\end{figure*}

	\textbf{Reward ($\mathcal{R}$):} Reward ($\mathcal{R}$) evaluates the decision made by the policy network to help the agent for better performance.  Here, if the agent chooses $STOP$, the reward will be 0, if the agent chooses any relation which will not change the meta-path set, to punish this situation,  the reward will be -1. Otherwise, we define the reward as the improvement of the performance. The formulation of reward can be represented as follows:
	\begin{displaymath}
		\label{eq: reward}
		R(s_i, a_i)= N(s_i, a_i) - N(s_{i-1}, a_{i-1}),
	\end{displaymath}
	where $N(s_i, a_i)$ is the performance of recommendation task at step $i$. Here, we adopt the NDCG@10 as the performance metric, but other metrics could also be used. After the policy network predicts the relation, we extend the current meta-path set and put them as the input of the downstream recommender. Then we train the recommender slightly using our proposed early stopping approximation technique (details in Section~\ref{subsec: rms_training}) and evaluate the model on the validation set to get the performance, then the reward is calculated to update the policy network.

\subsection{Model Training and Optimization}
\label{subsec: rms_training}

\textbf{Early Stopping Approximation.} For each meta-path set, if we train the recommender to convergence and then test the model to calculate rewards, the whole training time of the RL agent will be extremely long. Fortunately, inspired by the hyperparameter optimization techniques~\cite{ICML20StandleyZCGMS20, IJCAI15DomhanSH15, JMLR17LiJDRT17}, we find the performance of a model after training several epochs is highly correlated to the performance when the model converges, which means we can compare the quality of different meta-path sets on the lightly trained models. When we train the RL agent, we only lightly train the recommender and then evaluate the quality of generated meta-path sets. This strategy reduces the training time by more than \textbf{100 times}. However, we admit the correlation between lightly trained model performance and final model performance is not impeccable, it does have an accuracy penalty and the selected meta-paths are not necessarily optimal, but our experiments show that it is still very effective and can find high-quality meta-paths. 

\textbf{Training and Inference.} The whole procedure of training can be reviewed in Figure \ref{fig:overview}. The main steps are: i) get the current state based on the current meta-path set; ii) generate the next action (relation) based on the current state; iii) extend the current meta-path set and get the next state; iv) put the new meta-path set into the recommender and get the reward to update the RL agent. During the inference process, we start with the initial meta-path set and let the fully-trained RL agent generate relations and extend this set until the RL agent outputs the $STOP$ action or the current step exceeds the maximal step limit $I$.

\textbf{Optimization.} 
In our problem, States ($\mathcal{S}$) are in a continuous space, and Actions ($\mathcal{A}$) are in a discrete space. Therefore, we follow the steps of  Deep Q-Network~\cite{mnih2015human} (DQN), a typical RL algorithm, although it can be substituted with other RL algorithms. 
	Suppose that we have a policy network $Q$ which is used to evaluate an action on a state: $\mathcal{S} \times \mathcal{A} \rightarrow \mathbb{R}$. We can directly construct a policy $\pi$ that maximizes the reward:

	\begin{displaymath}
	\label{eq: argmax}
	\pi(s)=\mathop{argmax}\limits_{a} Q(s, a).
	\end{displaymath}
	
	The Bellman equation~\cite{mnih2015human} is used to update $Q$ as follows,

	\begin{displaymath}
		\label{eq: bellman}
		Q(s_i, a_i)=R(s_i, a_i)+\gamma Q\left(s_{i + 1} , \pi\left(s_{i+1}\right)\right),
	\end{displaymath}
	where $s_{i+1}$ is the next state and $a_{i+1}$ is the next action. In our algorithm, we utilize multi-layer perceptron (MLP) as the policy network $Q$ since it can approximate any function.

While training the agent, we first calculate the temporal difference error $\delta$ as below:
\begin{displaymath}
	\delta=Q(s, a)-\left(r+\gamma \max _{a} Q\left(s^{\prime}, a\right)\right).
\end{displaymath}

In order to minimize the error $\delta$, we use Huber loss~\cite{huber} function as shown below:
\begin{displaymath}
\mathcal{L}=\frac{1}{|B|} \sum_{\left(s, a, s^{\prime}, r\right) \in B} \mathcal{L}(\delta),
\end{displaymath}
where $\quad \mathcal{L}(\delta)=\left\{\begin{array}{ll}\frac{1}{2} \delta^{2} & |\delta| \leq 1 \\ |\delta|-\frac{1}{2} & \text { otherwise }\end{array}\right.$ and $B$ is randomly sampled history data.



\section{Our Recommender}
\label{sec: hrec}


Figure~\ref{fig:recommender} shows the architecture of our recommender \rec, containing two attention layers that learn the importance of different node neighbors and the different meta-paths, respectively. Here, we require a set of user meta-paths and a set of item meta-paths as input. User (Item) meta-paths means the meta-paths that start and end with a user (item) type node.

First, we will generate a meta-path reachable subgraph for each meta-path. The user (item) meta-path subgraphs will contain only user (item) type nodes, and the nodes are connected if they are meta-path neighbors. Then we perform a GAT~\cite{GATvelivckovic2018graph} layer to each subgraph to get the inter-meta-path information.  
The GAT~\cite{GATvelivckovic2018graph} layer can learn to differentiate the importance of different meta-path neighbors and aggregate the weighted information from them. Then the aggregated embeddings will be put into the next attention layer.

Since different meta-paths may play a different role in a recommendation task, we also adopt an attention mechanism to learn to assign the weight for different meta-paths and fuse the embeddings from each meta-path to get better node representations.
Suppose that we have $X$ meta-paths $\{\phi_1, \dots , \phi_X\}$. After GAT layers, we get $X$ groups of node embeddings denoted as $\{H_1, \dots , H_X\}$. 
To get the importance of each meta-path, we transform the embeddings via a multi-layer perceptron (MLP) and multiply with a query vector $\vec q_{\phi}^T$. Then get the average of all the meta-path-specific node embeddings. The equation is shown as follows:
\begin{displaymath}
	\label{eq: m_att}
	w^{\phi_x} =\frac{1}{|\mathcal{V}_x|}  \sum_{i \in \mathcal{V}_x}  \vec q_{\phi_x}^{T}  \cdot  \sigma( MLP(h_i^{\phi_x}) ), \  h_i^{\phi_x} \in H_x.
\end{displaymath} 

\begin{displaymath}
	\label{eq: m_norm}
	\beta^{\phi_x}=\frac{\exp(w^{\phi_x})}{\sum_{x = 1}^{X}\exp(w^{\phi_x})}.
\end{displaymath}

Here, $\beta^{\phi_x}$ denotes the normalized importance of meta-path $\phi_x$.  $\sigma$ is the activation function and $\mathcal{V}_x$ is the node set of meta-path $\phi_x$.
Lastly, we aggregate the embeddings from each meta-path to get the final node embeddings $H$ according to their weights by:
\vspace{-0.3em}
\begin{displaymath}
	\label{eq: m_agg}
	H =\sum_{x = 1}^{X} \beta^{\phi_x} \cdot H_x.
\end{displaymath}



\textbf{Recommendation.} After we get the neighbor-aggregated embeddings for users and items, we calculate the inner product of them to predict the scores that how much an item matches a user:

 \begin{displaymath}
    	\label{eq: inner}
    	\hat{y}(u, i) = h_u^T h_i.
 \end{displaymath}               

To train our recommendation model, we adopt BPR loss ~\cite{RendleFGS09}, which makes the scores of observed user-item iterations larger than the scores of unobserved interactions:

 \begin{displaymath}
	\label{eq: rec_loss}
	\mathcal{L}_{rec} = \sum_{(u, i, j) \in \mathcal{O}}^{} -ln\ \sigma (\hat{y}(u, i) - \hat{y}(u, j)), 
\end{displaymath}      
where $\mathcal{O}=\left\{(u, i, j) \mid(u, i) \in \mathcal{R}^{+},(u, j) \in \mathcal{R}^{-}\right\}$ denotes the training set. $\mathcal{R}^{+}$ means the positive (interacted) user-item pairs and $\mathcal{R}^{-}$ denotes the negative (un-interacted) user-item pairs, $\sigma(\cdot)$ represents the sigmoid function.     
For the training steps, we adopt the embeddings learned by Matrix Factorization (MF) ~\cite{RendleFGS09} as the initial embeddings of user and item nodes.

\textbf{Training strategies.} 
As mentioned above, \rec \ leverages a graph neural network to learn the representations of users and items. Nevertheless, graphs in recommender systems are often large-scale, incurring huge memory consumption and computation burden during message passing for graph neural networks, which makes it impossible to be applied to recommender systems directly when the given graph is huge. Therefore, we design two strategies to improve the efficiency and effectiveness of the model. 
\begin{enumerate}
	\item In the real world, a graph is often scale-free, and some nodes on an HIN may have too many meta-path neighbors, resulting in a huge time and memory cost. Therefore, we employ the Neighbor Sampling (NS) technique, which will randomly sample a subset of meta-path neighbors to perform message passing each time. This technique allows us to avoid running out of memory and save a lot of time.
	
	\item Meta-path is an effective tool to select semantic-related nodes. However, if a node is connected to too many nodes in the graph via a meta-path, then this meta-path should be ineffective because it cannot distinguish which nodes are highly related to a particular node. Therefore, for each meta-path, we will limit the density of the generated subgraph. We will discard the meta-path whose generated subgraph's density is larger than a given threshold $t$ (0 < $t$ < 1). This strategy further helps us filter out low-quality meta-paths.
\end{enumerate}

\textbf{Integrated with \rlmp.} According to the discussion above, \rec \ needs a set of user meta-paths and a set of item meta-paths as input. To cooperate with \rlmp, we train two RL agents to get the user meta-path set and item meta-path set separately. When we train the agent for the user meta-path set, we fix the item meta-path set as the initial set and vice versa. The RL agent will be updated by the feedback from the recommender.




\section{Experiments} \label{sec:exp}

\begin{figure*} 
  \subfloat[Yelp]{
  \label{fig: schema1} 
  \includegraphics[width=0.25\linewidth]{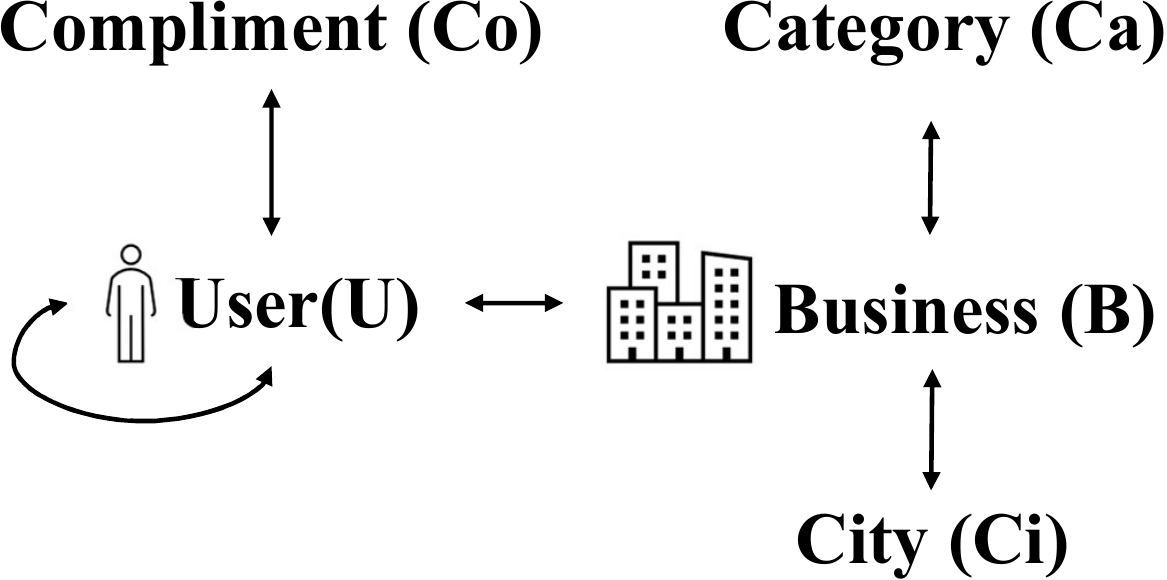}}
  \hspace{.2in} 
  \subfloat[Douban Movie]{
  \label{fig: schema2}
  \includegraphics[width=0.33\linewidth]{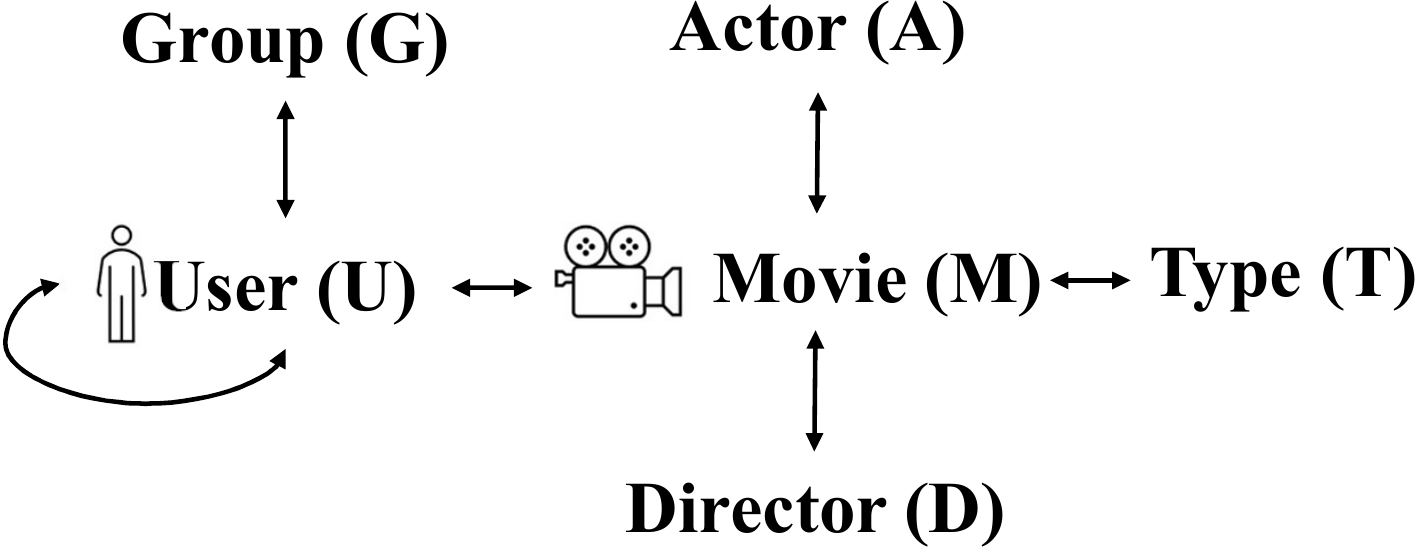}}
  \hspace{.1in} 
  \subfloat[TCL]{
  \label{fig: schema3}
  \includegraphics[width=0.3\linewidth]{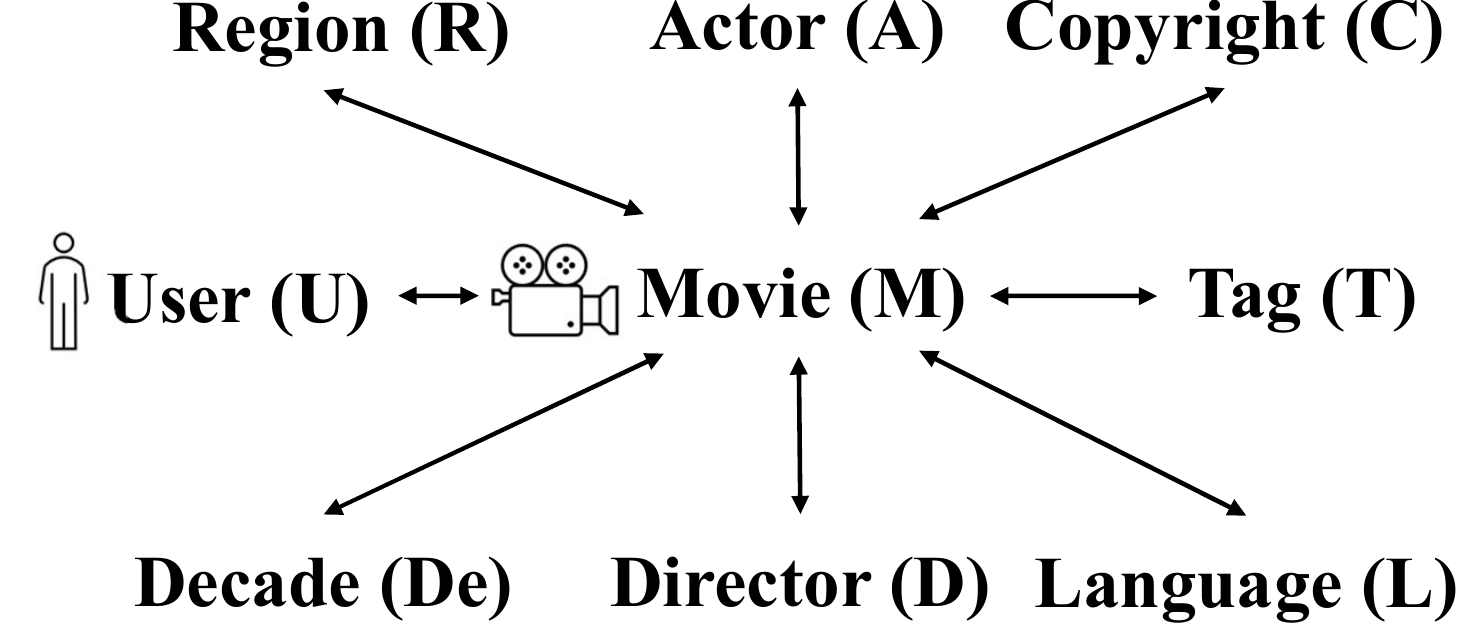}}
  \caption{Graph schema of three datasets}
  \label{fig: schema} 
  \vspace{-1em}
  \end{figure*}
  
\begin{table}[]
  \centering
  \begin{tabular}{|c|c|c|c|}
  \hline
              & \textbf{Yelp} & \textbf{Douban Movie} & \textbf{TCL} \\ \hline
  \#Nodes     & 31,092        & 37,595                & 122,284      \\ \hline
  \#Edges     & 976,276       & 3,429,882             & 1,084,154    \\ \hline
  \#Relations & 10            & 12                    & 16           \\ \hline
  \end{tabular}
  \caption{Statistics of the datasets}
  \label{tab: dataset}
  \vspace{-2em}
  \end{table}
  
We conduct experiments to answer these research questions. 

\begin{itemize}
    \item \textbf{RQ1:} How does \rlmp \ perform compared to other meta-path selection strategies?
    \item \textbf{RQ2:} How does the \rlmp -explored meta-path set perform compared to other meta-path sets?
    \item \textbf{RQ3:} How does \rlrec \ perform compared to other state-of-the-art recommendation models?
\end{itemize}

\subsection{Experiment Settings}


\subsubsection{Datasets.} We conduct our experiments on two commonly used public datasets \textbf{Yelp}\footnote{https://www.yelp.com/dataset/download} and \textbf{Douban Movie}\footnote{https://movie.douban.com/}, and a private industry dataset from \textbf{TCL}\footnote{https://www.tcl.com/}. The statistics and schema of the datasets are shown in Table \ref{tab: dataset} and Figure ~\ref{fig: schema}, respectively.

\subsubsection{Experiments Methods and Metrics.} To evaluate the performance of recommendation,  we adopt the leave-one-out evaluation~\cite{genetic2020han, peagnn2020han}. For each user, we will regard the interacted items as positive items and the remaining as negative items. We randomly select 499 negative samples for each positive item and rank the positive one among the total 500 items. Further, we adopt two common metrics~\cite{genetic2020han, KGAT19}, \textit{Hit Ratio at Rank k} (HR@k) and \textit{Normalized Discounted Cumulative Gain at Rank k} (NDCG@k) to evaluate our proposed models. 

\subsubsection{Baseline Meta-path Selection Strategies.}  
Since there exists no method that can derive different meta-paths for different MPRs,
to demonstrate the effectiveness of \rlmp \ integration on MPRs, we design two baseline meta-path selection strategies based on random and greedy selection. We also select a meta-path discovery method in graph mining area. 
\begin{itemize}
    \item \textbf{Random}: 
    We iteratively select a random meta-path set and perform training on the recommender using the selected meta-path set as input. We stop searching when the time limit is reached and select the meta-path set that achieves the best validation performance.
    \item \textbf{Greedy}: We initialize a meta-path set using the same process as in \rlmp, and in each iteration, we select a meta-path that achieves the best validation performance among the randomly picked meta-paths and add it to the meta-path set. Then train the recommender with the current meta-path set. Until the time limit is reached, we consider the final meta-path set as the current optimal meta-path set. 
    \item \textbf{MPDRL}~\cite{WanDPH20}: This is a meta-path discovery method based on the user-given node pairs in graph mining area which is proposed to gain insights from data. 
\end{itemize}

We run all of the meta-path selection strategies within the same time limit to ensure fairness in the meta-path selection process. Note that we adopt the early stopping approximation strategy on the recommender when we train the RL agent, so the RL agent can explore sufficient possible meta-path sets.   After the RL agent is well trained on the training set, we use the found meta-path set and train the recommender until the model converges.  Finally, we test the recommender on the test dataset.

\subsubsection{Baseline recommenders}
\textbf{Meta-path-based Recommenders.} To demonstrate the easy adaptation of \rlmp \ to MPRs, we first apply \rlmp \ to our recommender \rec \  and further apply it to two other representative state-of-the-art MPRs including HERec~\cite{herec2018shi} and MCRec~\cite{hu2018leveraging}. They need different forms of meta-paths as input.
\begin{itemize}
    \item \textbf{HERec}~\cite{herec2018shi}: 
    HERec needs two kinds of meta-path sets. One is the meta-paths start and end with a user type node and the other is the meta-paths start and end with an item type node.
    \item \textbf{MCRec}~\cite{hu2018leveraging}: 
    MCRec needs only one kind of meta-path set. The meta-paths should start with a user type node and end with an item type node.
\end{itemize}

\textbf{Non-meta-path-based Recommenders.} In addition to the aforementioned baseline models, we present recommendation performance achieved by other state-of-the-art models that are not based on the meta-paths, e.g., Matrix Factorization (MF)-based models (BPR~\cite{RendleFGS09}, NCF~\cite{he2017neural}), regularization-based models (CKE~\cite{ZhangYLXM16}, CFKG~\cite{ZhangCFKG}) and GNN-based models (GEMS~\cite{genetic2020han}, KGAT~\cite{KGAT19}).

\subsubsection{Implementation details.}
We implement all the models using Pytorch and DGL\footnote{http://dgl.ai/}. For the recommendation part of \rlrec,  we fix the embedding size to 64, the hidden layer size to 32, batch size to 90000. For the DQN of \rlmp, we adopt the implementation in ~\cite{rlcard2019zha} and conduct a 3-layer MLP with (32, 64, 32) hidden units for Q function. The learning rate and memory buffer size are set to 0.001 and 10000, respectively. We also set the maximal step limit to 4 for \rlrec, 3 for HERec and 5 for MCRec. We run the baseline recommenders using the default settings in their released code and original reports.


\subsection{Meta-path Selection Method Study (RQ1)}

\begin{table*}[]
  \centering
  \begin{tabular}{|cc|cc|cc|cc|}
  \hline
                                               &          & \multicolumn{2}{c|}{Yelp}         & \multicolumn{2}{c|}{Douban Movie} & \multicolumn{2}{c|}{TCL}          \\ \hline
  \multicolumn{1}{|c|}{Algorithm}              & Strategy & HR@3            & NDCG@10         & HR@3            & NDCG@10         & HR@3            & NDCG@10         \\ \hline
  \multicolumn{1}{|c|}{\multirow{4}{*}{\rec}}  & RMS      & \textbf{0.1484} & \textbf{0.1740} & \textbf{0.2131} & \textbf{0.2400} & \textbf{0.3079} & \textbf{0.3230} \\
  \multicolumn{1}{|c|}{}                       & Greedy   & 0.1381          & 0.1633          & 0.1800          & 0.2076          & 0.2895          & 0.3058          \\
  \multicolumn{1}{|c|}{}                       & Random   & 0.1181          & 0.1449          & 0.1946          & 0.2221          & 0.2875          & 0.3035          \\
  \multicolumn{1}{|c|}{}                       & MPDRL    & 0.1344          & 0.1589          & 0.1757          & 0.2041          & 0.2815          & 0.2993          \\ \hline
  \multicolumn{1}{|c|}{\multirow{4}{*}{HERec}} & RMS      & \textbf{0.0979} & \textbf{0.1215} & \textbf{0.1613} & \textbf{0.1984} & \textbf{0.2891} & \textbf{0.3082} \\
  \multicolumn{1}{|c|}{}                       & Greedy   & 0.0918          & 0.1166          & 0.1577          & 0.1960          & 0.2738          & 0.2920          \\
  \multicolumn{1}{|c|}{}                       & Random   & 0.0922          & 0.1167          & 0.1575          & 0.1945          & 0.2709          & 0.2878          \\
  \multicolumn{1}{|c|}{}                       & MPDRL    & 0.0915          & 0.1160          & 0.1570          & 0.1928          & 0.2713          & 0.2855          \\ \hline
  \multicolumn{1}{|c|}{\multirow{4}{*}{MCRec}} & RMS      & \textbf{0.1317} & \textbf{0.1540} & \textbf{0.1961} & \textbf{0.2236} & \textbf{0.1734} & \textbf{0.2075} \\
  \multicolumn{1}{|c|}{}                       & Greedy   & 0.1234          & 0.1504          & 0.1952          & 0.2204          & 0.1733          & 0.2070          \\
  \multicolumn{1}{|c|}{}                       & Random   & 0.1229          & 0.1451          & 0.1928          & 0.2193          & 0.1732          & 0.2066          \\
  \multicolumn{1}{|c|}{}                       & MPDRL    & 0.1223          & 0.1430          & 0.1919          & 0.2178          & 0.1694          & 0.2045          \\ \hline
  \end{tabular}
  \caption{Effect of different meta-path selection strategies}
  \label{tab:mpselect}
  \vspace{-2em}
  \end{table*}


  \begin{table*}[]
    \centering
    \begin{tabular}{|ccc|ccc|ccc|}
    \hline
    \multicolumn{3}{|c|}{Yelp}                                              & \multicolumn{3}{c|}{Douban Movie}                                      & \multicolumn{3}{c|}{TCL}                                               \\ \hline
    \multicolumn{1}{|c|}{Meta-path set} & HR@3            & NDCG@10         & \multicolumn{1}{c|}{Meta-path set} & HR@3            & NDCG@10         & \multicolumn{1}{c|}{Meta-path set} & HR@3            & NDCG@10         \\ \hline
    \multicolumn{1}{|c|}{RMS}           & \textbf{0.1484} & \textbf{0.1740} & \multicolumn{1}{c|}{RMS}           & \textbf{0.2131} & \textbf{0.2400} & \multicolumn{1}{c|}{RMS}           & \textbf{0.3079} & \textbf{0.3230} \\ \hline
    \multicolumn{1}{|c|}{- UBUU}        & 0.1441          & 0.1696          & \multicolumn{1}{c|}{- MAM}         & 0.2127          & 0.2381          & \multicolumn{1}{c|}{-UMU}          & 0.3057          & 0.3211          \\
    \multicolumn{1}{|c|}{- BCaB}        & 0.1425          & 0.167           & \multicolumn{1}{c|}{- MDMAM}       & 0.2104          & 0.2375          & \multicolumn{1}{c|}{-MDM}          & 0.3055          & 0.3213          \\
    \multicolumn{1}{|c|}{- BCiB}        & 0.1399          & 0.1657          & \multicolumn{1}{c|}{- MAMAM}       & 0.2103          & 0.2384          & \multicolumn{1}{c|}{-MLM}          & 0.3018          & 0.3185          \\
    \multicolumn{1}{|c|}{- BUB}         & 0.1352          & 0.1597          & \multicolumn{1}{c|}{- MUM}         & 0.2091          & 0.2360          & \multicolumn{1}{c|}{-MLMDM}        & 0.3012          & 0.3169          \\
    \multicolumn{1}{|c|}{}              &                 &                 & \multicolumn{1}{c|}{- MUMDM}       & 0.2083          & 0.2349          & \multicolumn{1}{c|}{-UMDMU}        & 0.2841          & 0.3005          \\
    \multicolumn{1}{|c|}{}              &                 &                 & \multicolumn{1}{c|}{- MDM}         & 0.2076          & 0.2357          & \multicolumn{1}{c|}{-MUM}          & 0.2800          & 0.2965          \\
    \multicolumn{1}{|c|}{}              &                 &                 & \multicolumn{1}{c|}{- UMDMU}       & 0.2075          & 0.2345          & \multicolumn{1}{c|}{-MUMDM}        & 0.2713          & 0.2884          \\
    \multicolumn{1}{|c|}{}              &                 &                 & \multicolumn{1}{c|}{- UMU}         & 0.2027          & 0.2308          & \multicolumn{1}{c|}{}              &                 &                 \\ \hline
    \multicolumn{1}{|c|}{+ BUBCaB}      & 0.1375          & 0.1636          & \multicolumn{1}{c|}{+ UMAMU}       & 0.2096          & 0.2358          & \multicolumn{1}{c|}{+MTM}          & 0.3060          & 0.3212          \\
    \multicolumn{1}{|c|}{+ UBCiBU}      & 0.1340          & 0.1602          & \multicolumn{1}{c|}{+ UMUU}        & 0.2064          & 0.2333          & \multicolumn{1}{c|}{+MDMDM}        & 0.3040          & 0.3200          \\ \hline
    \end{tabular}
    \caption{Effect of meta-path on \rlrec}
    \label{tab:mpeffect}
    \vspace{-1em}
    \end{table*}

\textbf{Effectiveness of \rlmp.} We compare \rlmp \ with self-design baselines over three MPRs.
Table \ref{tab:mpselect} shows the experimental results, and some observations are given as follows:
\begin{itemize}
    \item On all the datasets and models, \rlmp \  constantly  outperforms random and greedy strategies on all metrics. 
    This demonstrates that our method can find more effective meta-paths.
    \item For MCRec, the performance improvement is not as much as in \rec. We argue that the performance gain highly depends on the recommendation algorithms. Some models do not leverage meta-paths effectively so that their performance is not highly sensitive to the selected meta-paths. 
\end{itemize}

  \begin{figure} 
    \subfloat[Yelp]{
    \label{fig: rms_ndcg1} 
    \includegraphics[width=0.47\linewidth]{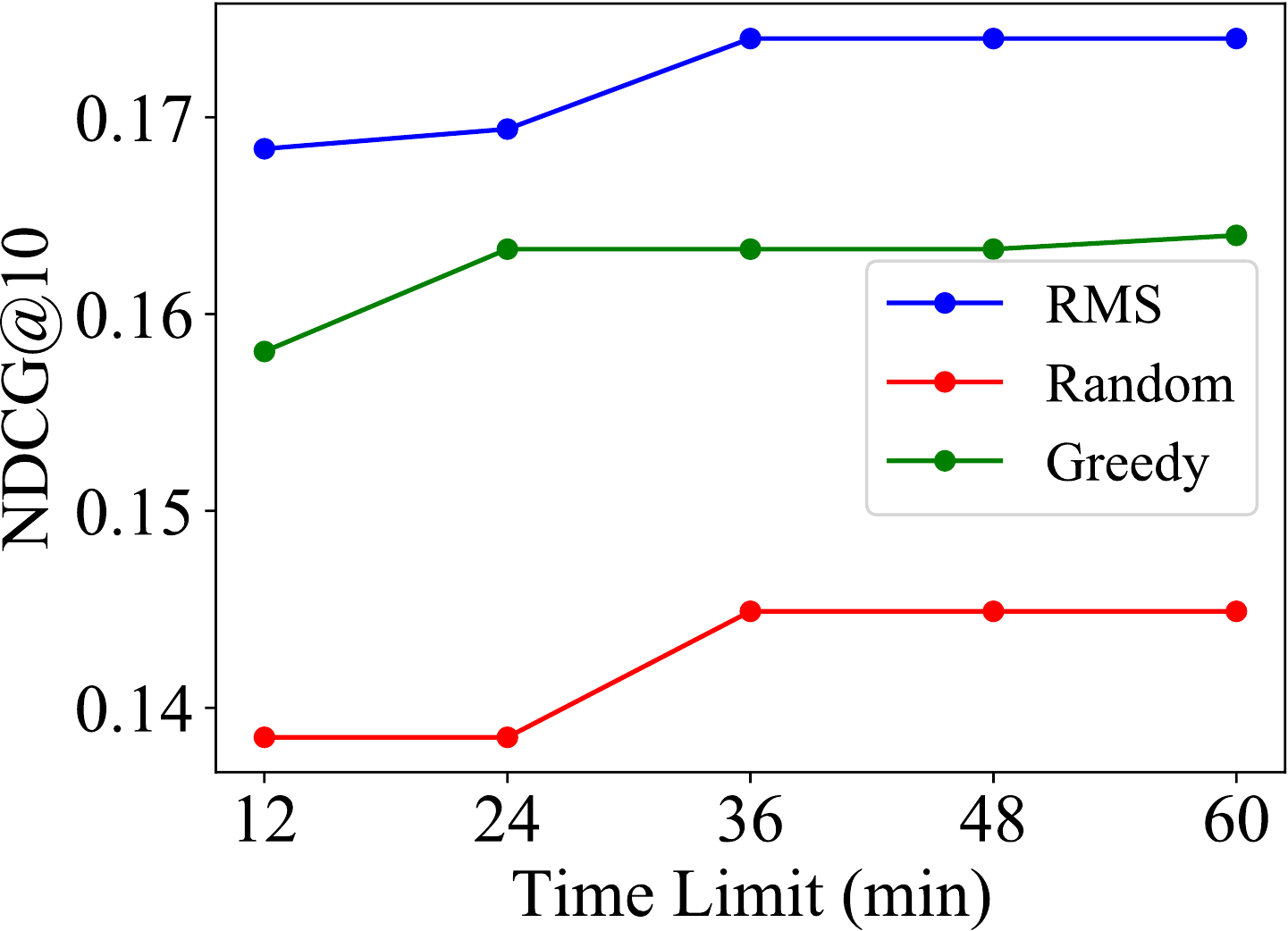}}
    \hspace{.1in} 
    \subfloat[Douban Movie]{
    \label{fig: rms_ndcg2}
    \includegraphics[width=0.47\linewidth]{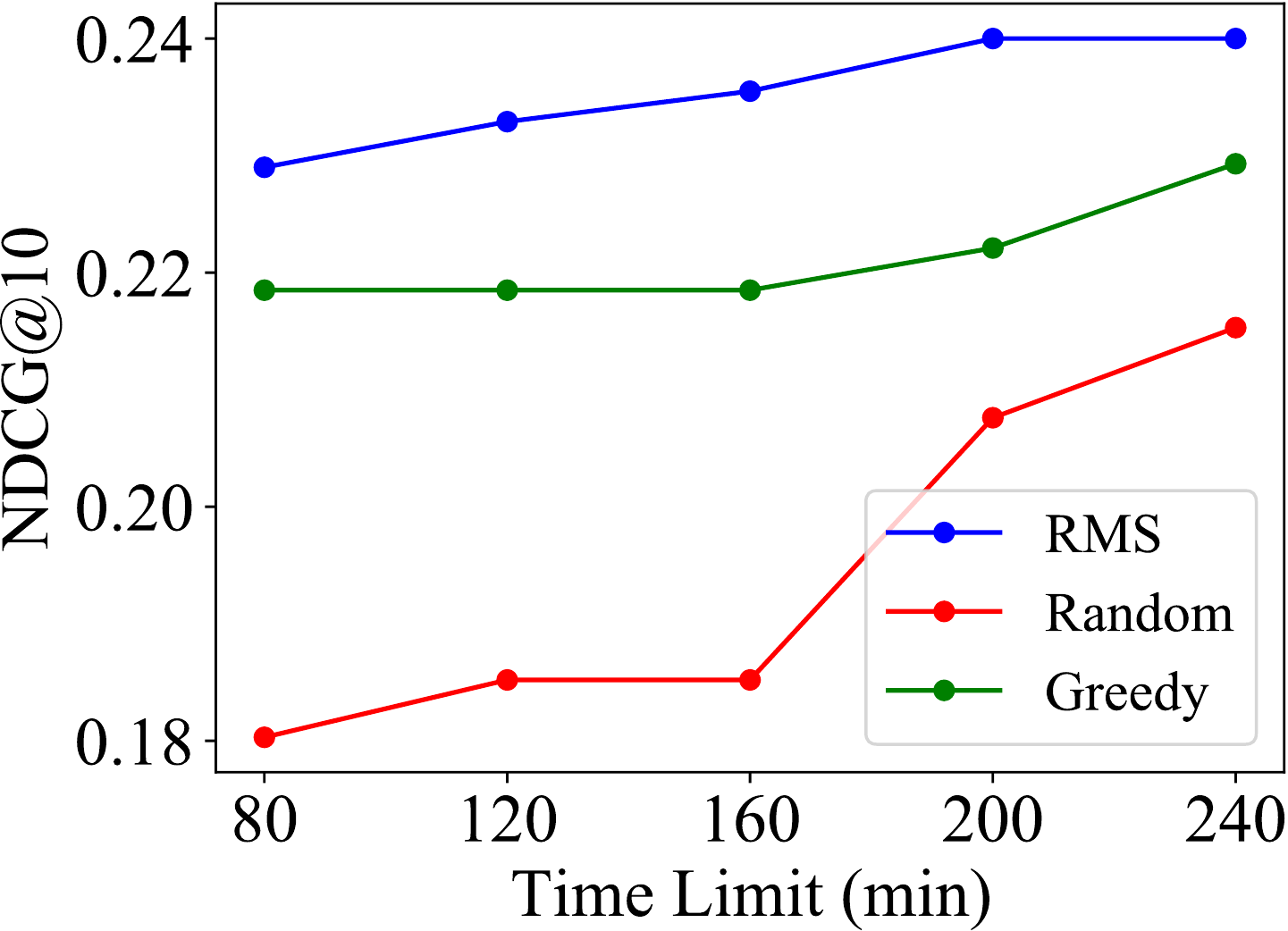}}
    \caption{NDCG@10 of each meta-path selection strategy over different time limits on \rec \ model}
    \label{fig: rms_ndcg} 
    \vspace{-1em}
  \end{figure}

\begin{figure} 
  \subfloat[Yelp]{
  \label{fig: yelp_time} 
  \includegraphics[width=0.43\linewidth]{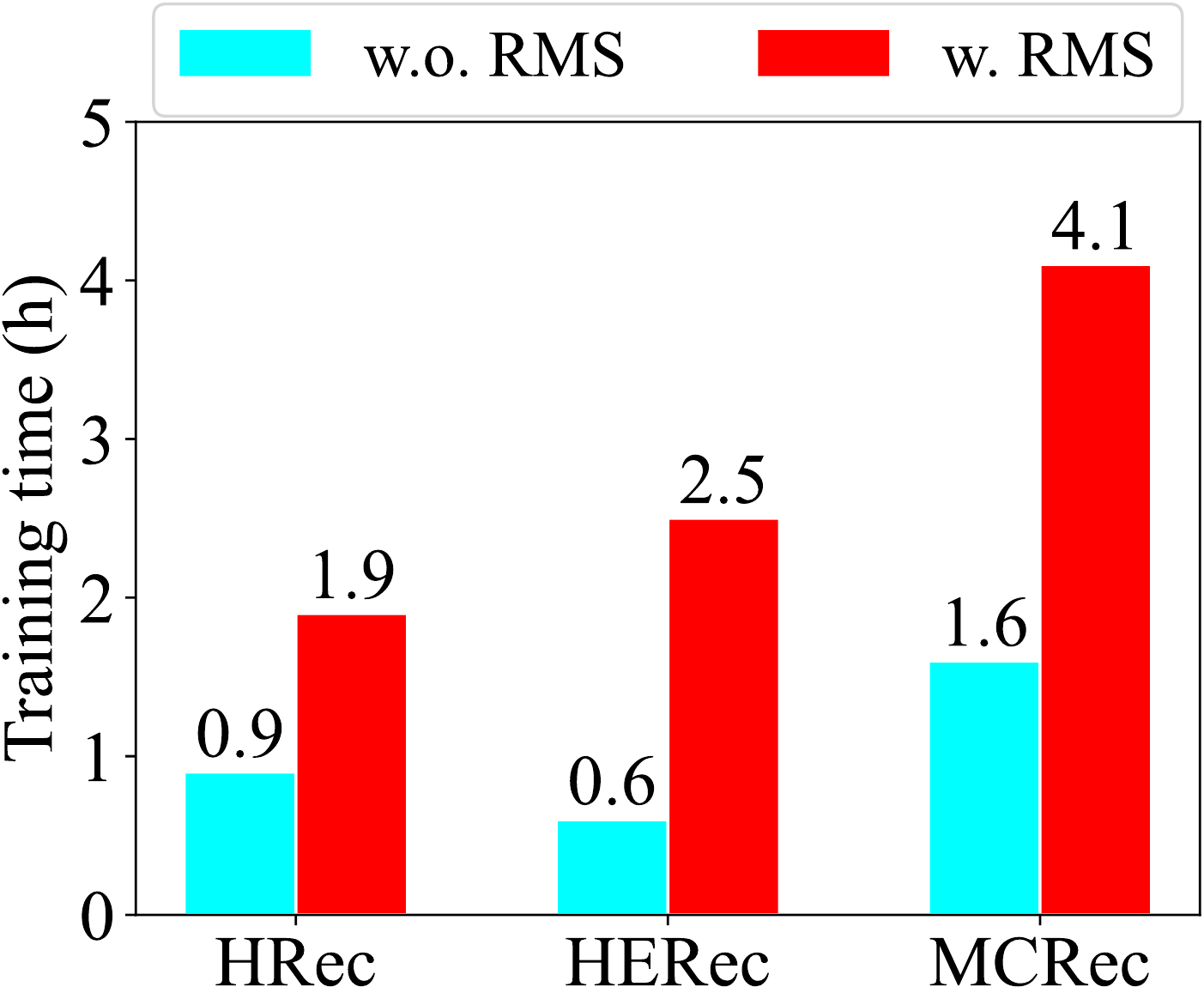}}
  \hspace{.1in} 
  \subfloat[Douban Movie]{
  \label{fig: douban_time}
  \includegraphics[width=0.44\linewidth]{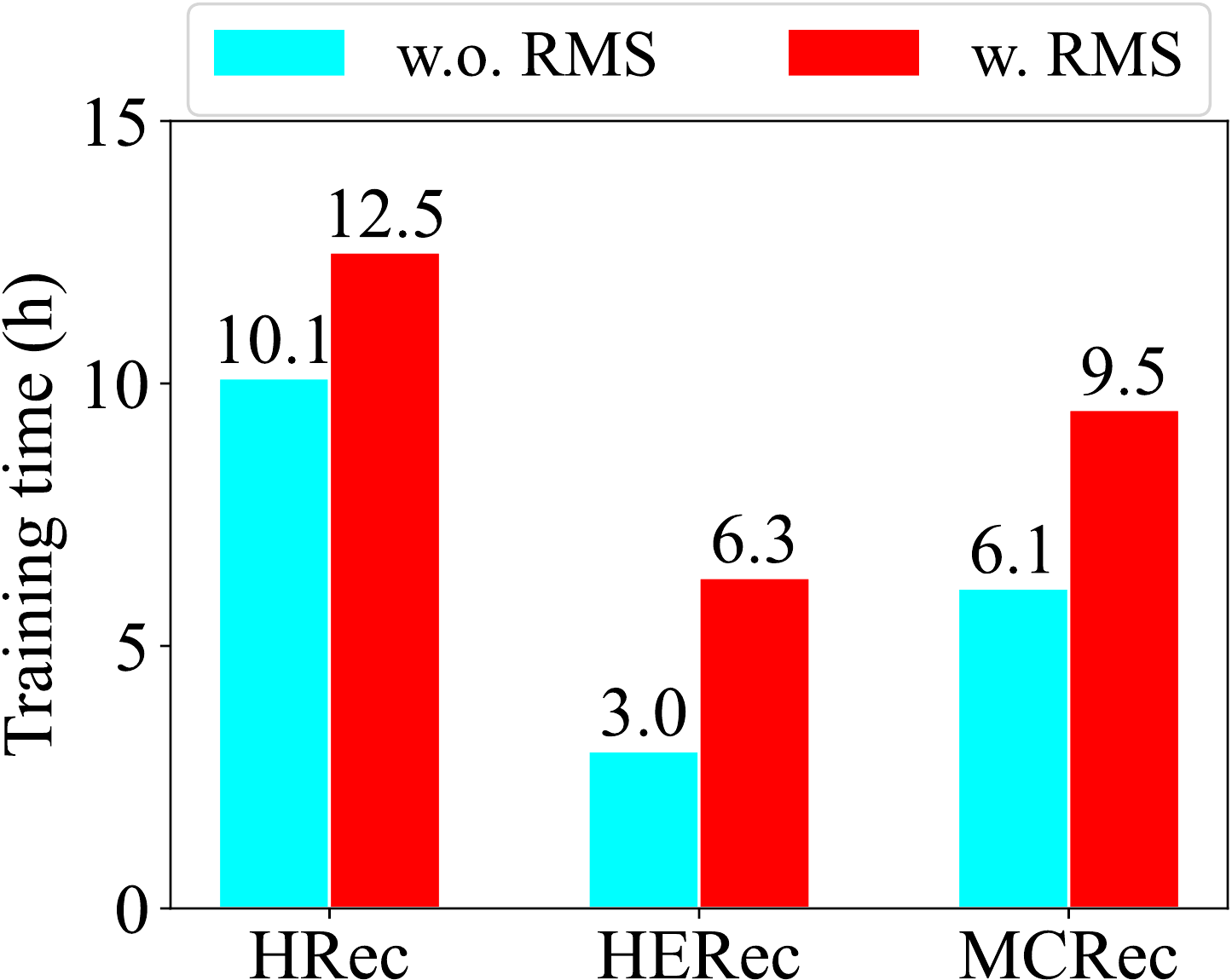}}
  \caption{Training time of each model with or without \rlmp}
  \label{fig: time} 
  \vspace{-1em}
  \end{figure}

\begin{figure}[]
	\centering
	\includegraphics[width=7cm]{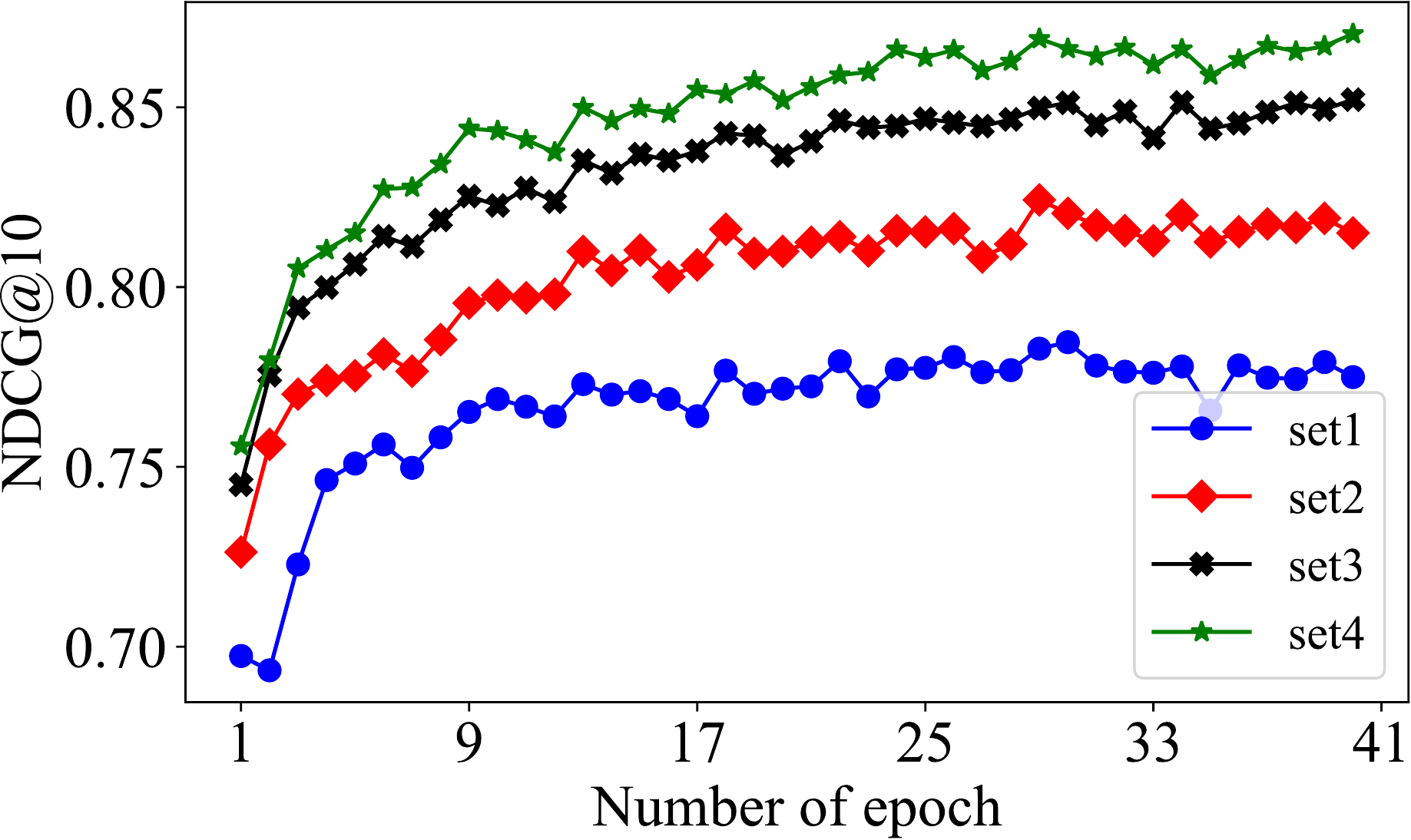}
	\caption{Case study for early stopping approximation: NDCG@10 of each epoch during training}
	\label{fig: one-epoch}
   \vspace{-2em}
\end{figure}

We also test the performance of each meta-path selection strategy over different time limit. Due to the space limitation, we only show the results on our \rec \ model on Yelp and Douban Movie dataset in Figure~\ref{fig: rms_ndcg}. It shows that \rlmp \ outperforms the baseline strategy in all situations. When the time limit is large, the NDCG@10 remains the same for \rlmp. This is because we stop training when the validation accuracy is decreasing, and the training process has finished early even though a lenient time limit is given. 

\textbf{Efficiency of \rlmp.} We also report the training time of each recommendation model on Yelp and Douban Movie datasets with (w.) or without (w.o.) \rlmp \ in Figure~\ref{fig: time}. We can find that even though using \rlmp \ takes an extra step to search meta-paths, the extra time is not very much.

\textbf{A case study of early stopping approximation.} During the training of RL model, for the efficiency purpose, we only train the recommender for a few epochs in our method, then evaluate the performance of the trained model on the small validation set. To prove this early stopping approximation approach is reasonable, we design an experiment on Yelp dataset. We manually designed four meta-path sets: (1) \textbf{Set1:} \{UU, BCiB\} (2) \textbf{Set2:} \{UU, UBU, BCiB\} (3) \textbf{Set3:} \{UBU, BUB\} (4) \textbf{Set4:} \{UBU, BUB, BCiB, BCaB\}.

We test each meta-path set on \rec \  on the validation set. As shown in Figure~\ref{fig: one-epoch}, we found that even though we lightly train the model (for one or a few epochs), the effectiveness of Set4 is the best, which is similar when the model converges. Though, we admit that this strategy may not guarantee the selected meta-paths are optimal, it greatly reduces the training time by more than \textbf{100 times}  and can also get decent results.

\subsection{Meta-path Sensitivity Analysis (RQ2)}

\begin{table}[]
  \centering
  \resizebox{\columnwidth}{!}{%
  \begin{tabular}{|c|cc|cc|cc|}
  \hline
      & \multicolumn{2}{c|}{Yelp}         & \multicolumn{2}{c|}{Douban Movie} & \multicolumn{2}{c|}{TCL}          \\ \hline
      & HR3             & NDCG10          & HR3             & NDCG10          & HR3             & NDCG10          \\ \hline
  RMS & \textbf{0.1484} & \textbf{0.1740} & \textbf{0.2131} & \textbf{0.2400} & \textbf{0.3079} & \textbf{0.3230} \\ \hline
  S1  & 0.1237          & 0.1514          & 0.1763          & 0.2015          & 0.2815          & 0.2993          \\ \hline
  S2  & 0.1420          & 0.1669          & 0.1987          & 0.2268          & 0.2995          & 0.3141          \\ \hline
  S3  & 0.1414          & 0.1696          & 0.1970          & 0.2255          & 0.3011          & 0.3178          \\ \hline
  \end{tabular}%
  }
  \caption{Performance of adding/removing more meta-paths}
  \label{tab:modifymore}
  \vspace{-2em}
  \end{table}

\begin{table*}[]
  \centering
  \begin{tabular}{|c|ccc|ccc|ccc|}
  \hline
         & \multicolumn{3}{c|}{Yelp}                           & \multicolumn{3}{c|}{Douban Movie}                   & \multicolumn{3}{c|}{TCL}                            \\ \cline{2-10} 
         & HR@1             & HR@3             & NDCG@10          & HR@1             & HR@3             & NDCG@10          & HR@1             & HR@3             & NDCG@10          \\ \hline
  \rlrec & \textbf{0.0648} & \textbf{0.1484} & \textbf{0.1740} & \textbf{0.0997} & \textbf{0.2131} & \textbf{0.2400} & \textbf{0.1654} & \textbf{0.3079} & \textbf{0.3230} \\ \hline
  BPR    & 0.0388          & 0.1025          & 0.1301          & 0.0529          & 0.1421          & 0.1768          & 0.1211          & 0.2704          & 0.2886          \\
  NCF    & 0.0514          & 0.1251          & 0.1522          & 0.0622          & 0.1605          & 0.1974          & 0.1279          & 0.2513          & 0.2673          \\ \hline
  CFKG   & 0.0572          & 0.1246          & 0.1477          & 0.0637          & 0.1650          & 0.2033          & 0.1602          & 0.2984          & 0.3145          \\
  CKE    & 0.0456          & 0.1092          & 0.1360           & 0.0612          & 0.1581          & 0.1947          & 0.1390          & 0.2818          & 0.3023          \\ \hline
  HERec  & 0.0548          & 0.1317          & 0.1540          & 0.0928          & 0.1961          & 0.2236          & 0.1484          & 0.2860          & 0.3055          \\
  MCRec  & 0.0389          & 0.0979          & 0.1215          & 0.0594          & 0.1613          & 0.1984          & 0.0602          & 0.1708          & 0.2056          \\ \hline
  GEMS   & 0.0100          & 0.0294          & 0.0408          & 0.0262          & 0.0591          & 0.0740          & 0.0344          & 0.0701          & 0.0923          \\
  KGAT   & 0.0415          & 0.1151          & 0.1388          & 0.0644          & 0.1669          & 0.2042          & 0.1454          & 0.2990          & 0.3162          \\ \hline
  \end{tabular}
  \caption{Overall Performance comparison of \rlrec}
  \label{tab:rec_perform}
  \vspace{-2em}
  \end{table*}

To illustrate the influence of meta-paths on recommendation performance and to demonstrate the effectiveness of \rlmp, we run \rlrec \ on all datasets. 

\textbf{Adding/removing one meta-path.} To test the meta-path sensitivity of \rlrec, we adopt a leave-one-out principle to check whether the performance changes if we remove/add one meta-path from/to the RMS-found meta-path set. 
Table~\ref{tab:mpeffect} shows the performance of \rlrec, \ where '-' means remove a meta-path, '+' means add a meta-path.
The findings on sensitivity analysis are listed below:

\begin{itemize}
    \item When we add/remove any meta-path to/from the meta-path set found by \rlmp , the performance will decrease, which proves that our approach can figure out the high-quality meta-paths.
    \item Performance deteriorates differently when we remove different meta-paths, demonstrating that different meta-paths have different impacts on performance. When we add another meta-path, the performance also drops, indicating that using as many meta-paths as possible does not necessarily lead to performance gains. 
\end{itemize}

\textbf{Adding/removing more meta-paths.} To explore the quality of other meta-path set. We also try to add/remove more meta-paths and test the performance. We just show some representative results\footnote{
For all datasets, S1 is the initial meta-path set that contain only two simplest meta-paths. S2 is a medium-size set and S3 is a large set. Here, Yelp S2 is also the handcrafting meta-path set used in ~\cite{herec2018shi}. For the other sets, we just randomly add or delete meta-paths from \rlmp-found set. We omit the specific meta-paths due to the space limit.} in Table~\ref{tab:modifymore} due to the space limit.  It is the performance of three handcrafted meta-path sets compared with the RMS-found set.
These results show if we remove/add more meta-paths, the performance may be even worse. Compared with the manually created meta-path sets, \rlmp~can also consistently improve the recommendation quality. Even compared with the meta-path set used in other paper (i.e., Yelp dataset S2~\cite{herec2018shi}), we can also get around 4.5\% performance gain in terms of HR@3.

\subsection{Recommendation Performance (RQ3)}

\textbf{Recommendation Effectiveness.} Table~\ref{tab:rec_perform} shows the overall performance of the algorithms in all datasets. For HERec and MCRec algorithm, instead of manually choosing a meta-path set, we use the best meta-path set found by \rlmp. The major observations are summarized as follows:

\begin{itemize}
    \item Our method outperforms all the baselines for all metrics on all datasets. 
    This  shows that the choice of meta-paths plays an important role in recommendation on HINs and our method can effectively leverage the semantic and structural information of the HINs and perform recommendation. 
    \item \rlrec \ outperforms two MF-based methods (BPR, NCF). This is because HINs have more entities besides users and items and contain more semantic information. \rlrec \ also has a better performance compared to two GNN-based methods (GEMS, KGAT) and two meta-path-based methods (HERec, MCRec), this shows using meta-paths and combining with GNNs is a better way to leverage HINs.
\end{itemize}

\begin{figure}[]
	\centering
	\includegraphics[width=7cm]{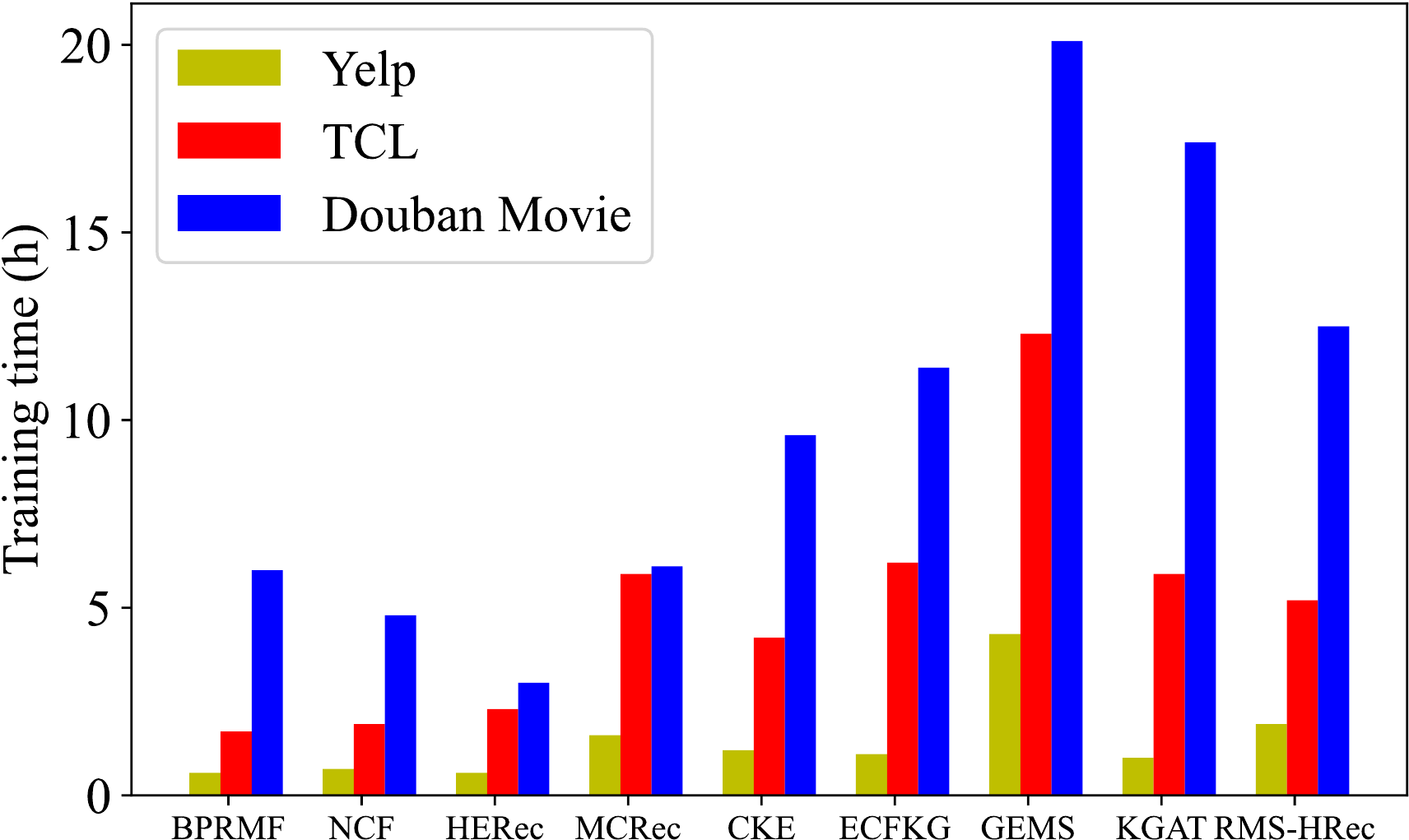}
	\caption{Training efficiency of each recommender}
	\label{fig: rec_time}
   \vspace{-1.3em}
\end{figure}

\textbf{Model Efficiency.} We report the training efficiency of \rlrec \ and all the baseline recommenders in Figure~\ref{fig: rec_time}. Our method needs to search meta-paths first, which will take some time. However, we find the training efficiency of our method is similar as or faster than other GNN-based methods (GEMS, KGAT) and not much slower than other methods. Therefore, we think the trade-off between accuracy and efficiency is reasonable.


\section{Related Work} \label{sec:rel}

\textbf{Meta-path-based recommenders.} 
Recently, Meta-paths are wildly used in a large volume of HIN-based recommendation models~\cite{han2019, WangWX00C19,DongCS17}. 
The MPRs can be divided into two categories according to the requirement of the input meta-path form. 
The first category~\cite{hu2018leveraging,FanZHSHML19} uses the meta-paths which start with the user type and end with the item type. 
This kind of algorithms can model the high-order connectivity and complex interactions between users and items. 
However, they rely on explicit path reachability and may not be reliable when path connections are sparse or noisy.
The second category~\cite{herec2018shi, peagnn2020han} needs two kinds of meta-paths: 1) the meta-paths that start and end with a user type, and 2) the meta-paths that start and end with an item type. 
They can fully mine latent structure features of users and items and are more resistant to sparse and noisy data, but may include too much information that is not relevant to the recommendation.
However, all of these methods suppose that meta-paths are already given, which is not realistic in real applications.

\textbf{Meta-path discovery on HINs.} Some works~\cite{ShiW14, MengCMSZ15, WanDPH20} attempt to discover meaningful meta-paths on HINs. 
GEMS~\cite{genetic2020han} adopts a genetic algorithm to find effective meta-structures (i.e., abstract graphs consisting of node and edge types) for recommendation and performs GCN~\cite{KipfW17} layers to fuse information from the found meta-structures. 
However, these meta-structures can only be used in their recommender since MPRs cannot handle meta-structures. Also, it is very time-consuming because of the colossal search space. 
RL-HGNN~\cite{abs-2010-13735} proposes a reinforcement learning (RL)-based method to find a meta-path for each node to learn its effective representations. However, they focus on producing powerful node embedding for each node using a found meta-path and enhancing the GNN model, rather than generating a set of format-specified meta-paths to support meta-path-based recommendation algorithms. Besides, it is not easy to adapt RL-HGNN to our scenario since all of their RL components are designed for obtaining better node embeddings, and our designs are completely different from theirs. In addition, it is not clear how their method can be adapted to generate meta-paths with specific forms.
According to the above discussion, even though some works are designed to find meta-paths, none of them can be directly adapted to benefit existing meta-path-based recommenders.

\section{Conclusion} \label{sec:con}

In this paper, we propose a reinforcement learning-based meta-path selection framework \rlmp \  to automatically find an effective meta-path set for recommendation. This allows recommender systems to better leverage information conveyed in meta-paths when making recommendations. Our experiments show that choices of meta-paths significantly affect the performance of meta-path-based recommenders and our proposed framework can effectively find high-quality meta-paths for any meta-path-based recommender. In addition, we design a new meta-path-based recommender, along with special training strategies, which better explore the potential of meta-paths and can achieve state-of-the-art performance.

\section*{Acknowledgements}
Reynold Cheng, Wentao Ning, and Nan Huo were supported by the HKU-TCL Joint Research Center for Artificial Intelligence (Project no. 200009430), the Innovation and Technology Commission of Hong Kong (ITF project MRP/029/18), the University of Hong Kong (Projects 104005858 and 10400599), and the Guangdong–Hong Kong-Macau Joint Laboratory Program 2020 (Project No: 2020B1212030009). Ben Kao was supported by HKU-TCL Joint Research Center for AI and Innovation and Technology Fund (Grant ITS/234/20).

\balance
\clearpage
\bibliographystyle{abbrv}
\bibliography{ref}


\end{document}